\documentclass[aps,prl,superscriptaddress,twocolumn,floatfix,showpacs]{revtex4}

\usepackage{amsmath}
\usepackage{graphicx}
\usepackage{bm}
\usepackage{mathrsfs}

\newcommand{\ppi}{\bm{\mathnormal{\Pi}}}

\begin{document}

\title{Electromagnetic wave collapse in a radiation background}

\author{Mattias Marklund}
\email{marklund@elmagn.chalmers.se}
\affiliation{Department of Electromagnetics, Chalmers University of
Technology, SE--412 96 G\"oteborg, Sweden}

\author{Gert Brodin}
\author{Lennart Stenflo}
\affiliation{Department of Physics, Ume{\aa} University, SE--901 87
Ume{\aa}, Sweden}

\date{\today}

\begin{abstract}
  The nonlinear interaction, due to quantum
  electrodynamical (QED) effects, between an electromagnetic pulse and
  a radiation background is investigated, by combining the methods of
  radiation hydrodynamics with the QED theory for photon-photon
  scattering. For the case of a single coherent electromagnetic pulse
  we obtain a Zakharov-like system, where the radiation pressure of
  the pulse acts as a driver of acoustic waves in the photon gas.
  For a sufficiently intense pulse and/or
  background energy density there is focusing and
  subsequent collapse of the pulse. The relevance of our results for
  various astrophysical applications are discussed.
\end{abstract}
\pacs{12.20.Ds, 95.30.Cq}

\maketitle

The theory for electromagnetism in vacuum can be roughly
divided into that for coherent and that for incoherent systems. In the
former case Maxwells equations are the explicit starting
point, and that theory leads to most of the wellknown applications.
In the latter case, the ensemble average of the electromagnetic field
is zero everywhere, and
the system can then essentially be treated as a gas
of ultra-relativistic particles \cite{Mihalas}. Most of
the theory for radiation gases cannot be considered as pure
``electromagnetism in vacuum'', since the interaction of
photons with other particles is one of the main features of the
theory. In the present
Letter, we will consider the \emph{self-interaction} 
between photons, due to quantum electrodynamical (QED) effects. 
One of the many interesting aspects
of such an analysis is the possibility of photon--photon
scattering, due to the interaction of light quanta with
\emph{virtual} electron--positron pairs
\cite{Heisenberg-Euler}. By integrating
out the virtual pairs, one obtains an effective field theory
for the photon--photon interaction, in terms of the
electromagnetic field variables
($F_{\mu\nu}$). The lowest order correction to Maxwell's vacuum
equations is then conveniently expressed by means of the
Heisenberg--Euler Lagrangian [see Eq.~(\ref{eq:lagrangian}) below].
In that way many papers on QED
photon--photon scattering have been written (e.g.,
\cite{Bialynicka-Birula,Boillat,Ding-Kaplan1,Ding-Kaplan2,Soljacic-Segev,%
Brodin-marklund-Stenflo,Brodin-etal,Boillat}
and references
therein).
Refs. \cite{Ding-Kaplan2,Soljacic-Segev,Brodin-marklund-Stenflo} 
concern techniques for 
laboratory detection of QED effects, 
involving second harmonic generation \cite{Ding-Kaplan2},
self-focusing \cite{Soljacic-Segev}, and nonlinear
wave mixing in cavities \cite{Brodin-marklund-Stenflo},
respectively. Other aspects of general 
theoretical interest that have been dealt with are, e.g., effects of
static magnetic background fields on higher harmonic generation, wave
propagation velocities in the QED vacuum and one-loop
corrections to the Heisenberg--Euler Lagrangian \cite{Ding-Kaplan1},
two-dimensional collapsing scenarios \cite{Brodin-etal}, and the
refractive properties of the QED vacuum \cite{Boillat}. However, the
above studies involve strictly coherent fields, or one-photon
phenomena. For instance, the non-trivial
propagation of photons in strong background
electromagnetic fields, due to effects of nonlinear electrodynamics,
has been considered in a number of papers (see, e.g., Refs.\
\cite{Bialynicka-Birula,Boillat} and
references therein). The main focus in these papers was on the
interesting effects of photon splitting and birefringence in
vacuum. However, Thoma \cite{Thoma} investigated the interaction of
photons with 
a photon gas, using the real-time formalism, and calculated the
corresponding change in the speed of light due to the Cosmic Microwave 
Background (CMB).

In the present letter we are going to combine methods from
radiation hydrodynamics \cite{Mihalas} with QED theory for
photon-photon scattering, in order to give a framework for
the interaction of coherent electromagnetic fields with a
radiation gas background. For the case of a single
coherent electromagnetic pulse we obtain a Zakharov-like
system \cite{Malomed-etal}, where the radiation pressure of the pulse
acts 
as a driver of acoustic waves in the photon gas. Similarly to
ordinary acoustic waves these waves are longitudinal and characterised
by variations in pressure, and thereby in energy density. 
The index
of refraction depends on the photon gas energy density, and thus
the excitation of the acoustic waves leads to a back-reaction on
the pulse. For a sufficiently intense pulse and/or
background energy density we may have focusing and
subsequent collapse of the pulse. Applications to coherent
pulses propagating in the present as well as in the early cosmic
radiation background are discussed, together with other astrophysical 
phenomena. While there are possible explanations for the
small CMB structures recently detected \cite{Boomerang}, we think that 
our mechanism below for CMB structure formation involving
photon--photon interaction should be of interest.

The nonlinear self-interaction of photons can be formulated in terms
of the Heisenberg--Euler Lagrangian \cite{Heisenberg-Euler}
\begin{equation}\label{eq:lagrangian}
  L = \varepsilon_0F +
  \kappa\varepsilon_0^2\left[4F^2 + 7G^2 \right],
\end{equation}
where $F = (E^2 - c^2B^2)/2$ and $G = c\bm{E}\cdot\bm{B}$.
Here $\kappa \equiv 2\alpha^2\hbar^3/45m_e^4c^5 \approx 1.63\times
10^{-30}\, \mathrm{m}\mathrm{s}^{2}/\mathrm{kg}$, $\alpha$ is
the fine-structure
constant, $\hbar$ is the Planck constant, $m_e$ the electron mass, and
$c$ the velocity of light in vacuum (for higher
order corrections, see, e.g., expression (26) of Ref.\
\cite{Bialynicka-Birula}). The Lagrangian (\ref{eq:lagrangian}) is
valid as long as there is no pair creation and the field strength is
smaller than the critical field, i.e., 
\begin{equation}
  \omega \ll m_ec^2/\hbar \,\,\text{ and }\,\, |\bm{E}| \ll
  E_{\text{crit}} \equiv 
  m_ec^2/e\lambda_c   
  \label{eq:constraint} 
\end{equation}
respectively. Here $e$ is the elementary charge, $\lambda_c$ is the
Compton wave length, and $E_{\text{crit}} \simeq
10^{18}\,\mathrm{V}/\mathrm{m}$.  

According to Ref. \cite{Bialynicka-Birula}, the dispersion relation
for a low energy photon in a background electromagnetic field can be
derived from the Lagrangian (\ref{eq:lagrangian}), with the result
(see also Ref.~\cite{Boillat} and references
therein)
\begin{equation}\label{eq:dispersionrelation}
  \omega(\bm{k}, \bm{E}, \bm{B}) = c|\bm{k}|\left( 1 -
  \tfrac{1}{2}\lambda|\bm{Q}|^2 \right) .
\end{equation}
where
\begin{eqnarray}
  |\bm{Q}|^2 &=& \varepsilon_0\left[ E^2 + c^2B^2
   -
   (\hat{\bm{k}}\cdot\bm{E})^2 -
   c^2(\hat{\bm{k}}\cdot\bm{B})^2 \right. \nonumber \\
   &&\qquad \left.  -
   2c\hat{\bm{k}}\cdot(\bm{E}\times\bm{B})\right] ,
\label{eq:Q2}
\end{eqnarray} 
and $\lambda = \lambda_{\pm}$, where $\lambda_+ = 14\kappa$ and
$\lambda_-
= 8\kappa$ for the two different polarisation states of the
photon. Furthermore, $\hat{\bm{k}} \equiv \bm{k}/k$.
The approximation $\lambda|\bm{Q}|^2 \ll 1$ has been used.
The background electric and magnetic fields are
denoted by $\bm{E}$ and $\bm{B}$, respectively.

We will below study two scenarios: (i) A plane wave pulse
propagating on a background consisting of a radiation gas in
equilibrium, and (ii) a radiation gas affected by an electromagnetic
(EM) pulse propagating through the gas. 
For case (i), the relations $
  (\hat{\bm{k}}_p\cdot\bm{E})^2 = \tfrac{1}{3}E^2$,
  $(\hat{\bm{k}}_p\cdot\bm{B})^2 = \tfrac{1}{3}B^2$, and
  $\bm{E}\times\bm{B} = 0$,
hold, where $\bm{k}_p$ is the pulse wave vector.
Hence, from (\ref{eq:Q2}) we obtain
\begin{equation}\label{eq:Q2gas}
  |\bm{Q}_{\text{gas}}|^2 = \tfrac{4}{3}\mathscr{E} ,
\end{equation}
where $\mathscr{E} = \varepsilon_0(E^2 + c^2B^2)/2$ is the energy
density of the radiation gas. 
For case (ii), we argue in a similar manner. To lowest order, the
directions
$\hat{\bm{k}}$ of the photons in the gas are
approximately random, and the EM pulse is a superposition
of uni-directional plane waves, such that $\bm{E} =
E_p\hat{\bm{e}}$, and $\bm{B} =
E_p\hat{\bm{k}}_p\times\hat{\bm{e}}/c$, where
$\hat{\bm{e}}$ is the unit electric vector.
Then (\ref{eq:Q2}) yields
\begin{equation}\label{eq:Q2pulse}
  |\bm{Q}_{\text{pulse}}|^2 =
   \tfrac{4}{3}\varepsilon_0|E_p|^2 .
\end{equation}

With the relation (\ref{eq:Q2gas}), and using standard methods for
slowly varying envelopes \cite{Hasegawa}, we then derive the
dynamical equation for a pulse on a photon gas background
\begin{equation}\label{eq:nlse}
  i\left( \frac{\partial}{\partial t} +
  c\hat{\bm{k}}_p\cdot\nabla
  \right)E_p +
  \frac{c}{2k_p}\nabla^2_{\perp}E_p +
  \frac{2}{3}\lambda ck_p\mathscr{E}E_p = 0 ,
\end{equation}
where $\nabla^2_{\perp} = \nabla^2 - (\hat{\bm{k}}_p\cdot\nabla)^2$,
noting that the \emph{self-interaction} of the pulse vanishes.

For a dispersion relation $\omega = ck/R(\bm{r}, t)$,
where $R$ is the refractive index, we have the
Hamiltonian ray equations
\begin{subequations}
\begin{eqnarray}
  \dot{\bm{r}} &=& \frac{\partial\omega}{\partial\bm{k}} =
  \frac{c}{R}\,\hat{\bm{k}}  ,
  \label{eq:groupvelocity} \\
  \dot{\bm{k}} &=& -\nabla\omega = \frac{\omega}{R}\nabla R ,
\label{eq:kdot}
\end{eqnarray}
\end{subequations}
where $\dot{\bm{r}}$ denotes the group velocity of the photon,
$\dot{\bm{k}}$ the force on a photon, and the dot denotes time
derivative.

The equation for the collective interaction of photons can
then be formulated as~\cite{Mendonca}
\begin{equation}\label{eq:kinetic}
  \frac{\partial N(\bm{k}, \bm{r}, t)}{\partial t} +
  \nabla\cdot(\dot{\bm{r}} N(\bm{k}, \bm{r}, t)) +
  \frac{\partial}{\partial\bm{k}}\cdot(\dot{\bm{k}}N(\bm{k}, \bm{r},
  t)) = 0 .
\end{equation}
where the distribution function has been normalised such that the
number density is $n(\bm{r}, t) = \int N(\bm{k}, \bm{r},t)\,d\bm{k}$.

For a general function $f(\bm{k}, \bm{r}, t)$, the moment equation is 
\begin{eqnarray}
  && \frac{\partial}{\partial t}(n\,\langle f\rangle) +
  \nabla\cdot \left( n\, \langle\dot{\bm{r}}f\rangle\right)
  \nonumber \\
  && \qquad
  = n\left[\,\, \left\langle{\frac{\partial f}{\partial t}}\right\rangle
  + \langle{\dot{\bm{r}}\cdot\nabla f}\rangle
  + \left\langle{\dot{\bm{k}}\cdot\frac{\partial
  f}{\partial\bm{k}}}\right\rangle 
  \right],
\label{eq:moment}
\end{eqnarray}
where $\langle{f}\rangle \equiv n(\bm{r}, t)^{-1}\int f N
  \,d\bm{k}$. 
Choosing $f = \hbar\omega$, we obtain the energy conservation
equation
\begin{subequations}
\begin{equation}\label{eq:energy}
  \frac{\partial\mathscr{E}}{\partial t} +
  \nabla\cdot\left( \mathscr{E}\bm{u} + \bm{q} \right) =
  -\frac{\mathscr{E}}{R}\frac{\partial R}{\partial
  t} ,
\end{equation}
where $\mathscr{E}(\bm{r}, t) = n\langle{\hbar\omega}\rangle$
is the energy density, and $\bm{q}(\bm{r}, t) =
n\,\langle{\hbar\omega\bm{w}}\rangle$
the energy flux.
Here we have made the split $\dot{\bm{r}} = \bm{u} + \bm{w}$, where
$\langle{\bm{w}}\rangle = 0$. Thus
$\bm{w}$ represents the random velocity of the photons. 
With $f = \hbar\bm{k}$, we obtain the momentum conservation
equation
\begin{eqnarray}\label{eq:momentum}
  \frac{\partial{\ppi}}{\partial t} + \nabla\cdot\Big[
  \bm{u}\otimes{\ppi} + \bm{\mathsf{P}} \Big] =
  \frac{\mathscr{E}}{R}\nabla R ,
\end{eqnarray}
\label{eq:comoving}
\end{subequations}
where ${\ppi} = n\,\langle{\hbar\bm{k}}\rangle$
is the momentum density, and
$\bm{\mathsf{P}} = n\,\langle{\bm{w}\otimes(\hbar\bm{k})}\rangle$
is the pressure tensor. It follows immediately from
the definition of the pressure tensor that the trace 
satisfies $\mathrm{Tr}\,\bm{\mathsf{P}} =
  n\, \langle{\hbar k\bm{w}\cdot\hat{\bm{k}}}\rangle
  =
  nR\,\langle{\hbar\omega\bm{w}\cdot\hat{\bm{k}}}\rangle/c$.
For an observer comoving with the fluid, i.e., a system in
which $(\bm{u})_0 = 0$ (the $0$ denoting the comoving system),
Eq.~(\ref{eq:groupvelocity}) shows that
$(\bm{w}\cdot\hat{\bm{k}})_0 = (R)_0/c$, so that the trace of the
pressure
tensor
in the comoving system becomes $(\mathrm{Tr}\,\bm{\mathsf{P}})_0 =
(\mathscr{E})_0$.
For an isotropic distribution function, the pressure can be written as
$P = \mathrm{Tr}\,\bm{\mathsf{P}}/3$, satisfying the equation of state
$P = \mathscr{E}/3$. 

The system of equations (\ref{eq:energy}) and (\ref{eq:momentum})
needs closure, and this can be obtained by choosing an equation of
state. Furthermore, we still have some gauge freedom since
we have not specified the frame. Two choices prevail in the
literature, (i) the energy frame,
in which $\bm{q} = 0$, and (ii) the particle frame, where $\bm{u}
= 0$. We will here adopt a particle frame, and furthermore choose the
equation of state to be $\mathsf{P}_{ij} =
\delta_{ij}\mathscr{E}/3$. The relation between the heat flow and the
momentum density is straightforward to derive. The result is
$\bm{q} = c^2{\ppi}/R^2$. 
From the dispersion relation (\ref{eq:dispersionrelation}) and
Eq.~(\ref{eq:Q2pulse}), we have
$R \approx 1 + \tfrac{2}{3}\lambda\varepsilon_0|E_p|^2$,
from which, through Eqs.~(\ref{eq:comoving}), we obtain
\begin{subequations}
\begin{eqnarray}
  \frac{\partial\mathscr{E}}{\partial t} +
  c^2\nabla\cdot{\ppi} &=& \frac{2}{3}\lambda\varepsilon_0\Bigg[ -
  \mathscr{E}%
  \frac{\partial|E_p|^2}{\partial t} \nonumber \\
  &&\qquad\quad +
  c^2\nabla\cdot\left(
  {\ppi}|E_p|^2 \right) \Bigg] ,
\label{eq:energyeq}
\end{eqnarray}
and
\begin{eqnarray}
  \frac{\partial{\ppi}}{\partial t} + \frac{1}{3}\nabla\mathscr{E}
  =  \frac{2}{3}\lambda\varepsilon_0
  \mathscr{E}\nabla|E_p|^2  .
\label{eq:momentumeq}
\end{eqnarray}
\label{eq:fluidequations}
\end{subequations}

We now introduce $\mathscr{E} = \mathscr{E}_0 + \delta\mathscr{E}$,
where $\mathscr{E}_{0}$ 
($\gg \delta \mathscr{E}$) is the unperturbed energy density in the
absence of the electromagnetic pulse, noting that for the unperturbed
state $\ppi = \bm{0}$. Taking the divergence of (\ref{eq:momentumeq})
and eliminating $\ppi$ using (\ref{eq:energyeq}), we obtain
\begin{subequations}
\begin{equation}
\frac{\partial ^{2}\delta \mathscr{E}}{\partial
t^{2}}-\frac{c^{2}}{3}\nabla
^{2}\delta \mathscr{E} = -\frac{2\lambda \varepsilon
_{0}\mathscr{E}_{0}}{3}%
\left[ c^2\nabla ^{2}\left| E_p \right| ^{2}+\frac{\partial
    ^{2}\left| E_p
\right| ^{2}}{\partial t^{2}}\right]   \label{acoustic-op}
\end{equation}
describing acoustic waves in a
photon gas driven by a coherent electromagnetic pulse.
We note that the set of equations (\ref{acoustic-op}) and
(\ref{eq:nlse}) 
resembles the wellknown Zakharov system, describing coupled Langmuir
and ion-acoustic waves in plasmas \cite{Malomed-etal}.  
Transforming to a system moving with the pulse $\tau
=t$, $\xi =x-ct$ (letting the pulse wave-vector be $\bm{k}_p =
k_p\hat{\bm{x}}$), we have 
\begin{equation}
\frac{\partial^2\delta \mathscr{E}}{\partial\xi^2} - %
\frac{1}{2}\nabla_{\bot}^2\delta \mathscr{E} = - \lambda
\varepsilon_0\mathscr{E}_0\left[ \nabla _{\bot }^{2}\left|
  E_p
\right| ^{2}+2\frac{\partial ^{2}\left| E_p \right|
  ^{2}}{\partial \xi
^{2}}%
\right]
\label{Coomoving-acoustic}
\end{equation}
\end{subequations}
dropping the slow derivatives proportional to $\partial /\partial \tau
$.
For a general pulse geometry Eq. (\ref{Coomoving-acoustic}) requires
extensive analysis, but for specific forms of the pulse we can
significantly simplify the
expression for $\delta \mathscr{E}$. For a broad electromagnetic
pulse where $\nabla _{\bot }\ll \partial /\partial \xi $ we
thus integrate (\ref{acoustic-op}) to obtain 
$\delta \mathscr{E} = -2\lambda \varepsilon
_{0}\mathscr{E}_{0}|E_p|^2$, 
whereas for a long needle shaped pulse with $\partial /\partial \xi \ll
\nabla _{\bot }$ we instead have 
$\delta\mathscr{E} = 2\lambda \varepsilon_0\mathscr{E}_0|E_p|^2$. 
The unperturbed photon gas energy density $\mathscr{E}_0$
will give a frequency shift for the solution of Eq.~(\ref{eq:nlse}),
but this shift can be transformed away. The resulting equation is
\begin{equation}
  i\frac{\partial E_p}{\partial\tau} +
  \frac{c}{2k_p}\nabla_{\perp}^2E_p
  \mp \frac{4}{3}\lambda^2ck_p\varepsilon_0\mathscr{E}_0|E_p|^2E_p = 0 ,
\label{eq:transformednlse}
\end{equation}
where the minus (plus) refers to a broad 
(needle shaped) pulse. For a
broad pulse, wave collapse does not occur, but for
a needle shaped pulse, collapse will occur when the
nonlinear term 
dominates over the diffraction. Comparing the terms in
Eq.~(\ref{eq:transformednlse}), we can obtain a rough estimate of the
collapse criterion, namely
\begin{equation}
  k_p^2r_p^2|E_p|^2|E_{\mathrm{gas}}|^{2} >  {|E_{\mathrm{char}}|^{4}}
  \label{Collapse-cond} 
\end{equation}
which is consistent with Ref.~\cite{Desaix-Anderson-Lisak}. Here $r_p$
denotes the pulse width, 
$\mathscr{E}_0 = \varepsilon_0|E_{\mathrm{gas}}|^2$,
and $|E_{\mathrm{char}}| \equiv
(\varepsilon _{0}\kappa )^{-1/2}\approx 2.6\times
10^{20}\,\mathrm{V/m}$.
Although the collapse of electromagnetic pulses is
complicated \cite{collapse}, the qualitative features can be 
described as a divergence in the pulse   
energy density, while its width decreases to zero, \emph{in a finite
  time}. A rough estimate of the 
collapse properties can be obtained assuming cylindrical symmetry of the
pulse. Following \cite{Desaix-Anderson-Lisak}, the trial function 
$
  E_T(\tau, r) = A(\tau)\,\text{sech}[r/a(\tau)]\exp[ib(\tau)r^2]
$ 
is used together with Rayleigh--Ritz optimisation, in order to reduce
the problem to a differential equation for the width $a(\tau)$. The
field strength then satisfies $|A(\tau)|/|A(0)| = a(0)/a(\tau)$, and the
width behaves as $(a(\tau)/a(0))^2 - 1 \propto ( 1 -
a(0)^2|A(0)|^2/I_c)\tau^2$ where $I_c \simeq
0.5\times(k_p^2\lambda^2\varepsilon_0\mathscr{E}_0)^{-1}$ 
Thus, if $a(0)^2|A(0)|^2 > I_c$ the pulse will collapse to zero width
in a finite time. Of course, when the pulse intensity increases,
higher order effects will become important, possibly halting the
collapse. Furthermore, as the pulse width decreases, derivative
corrections to (\ref{eq:lagrangian}) become
important \cite{Mamaev-etal}, and thus change the collapse scenario.   

The
most intense pulses in our universe are the gamma-ray bursts
\cite{Piran}. Powers of the order $10^{45}\,\mathrm{W}$ in the gamma
range ($k_p \sim 10^{13}\,\mathrm{m}^{-1}$) means
that we may have $|E_p|k_p r_p \sim
10^{38}\,\mathrm{V}/\mathrm{m}$ \footnote{Note that the given values
  correspond to field strengths $E_p < E_{\text{crit}}$ provided the
  width of the bursts satisfy the moderate condition $r_p > 10^7 \,
  \mathrm{m}$.}.  In todays universe the energy 
density of the CMB is $2.6\times10^5\,\mathrm{eV}/\mathrm{m}^3$, which
correspond to $|E_{\mathrm{gas}}| \sim 7\times
10^{-2}\,\mathrm{V}/\mathrm{m}$. Thus the collapse condition
(\ref{Collapse-cond}) is (far from surprising) not fulfilled
for gamma-ray bursts propagating in the present microwave background.
However, we note that if there were mechanisms for generating equally
intense gamma pulses in the early universe at an age $t \lesssim
6\times 10^5 \,\mathrm{yr}$, such pulses should indeed have collapsed,
as for those times the energy density of the CMB obeyed
$|E_{\mathrm{gas}}| \gtrsim 2\times 10^4 \,\mathrm{V}/\mathrm{m}$
implying that (\ref{Collapse-cond}) is fulfilled.

Furthermore, we note that collapse may occur for much less intense
pulses than gamma-ray bursts, in case the propagation takes place in a
photon gas which is more energetic than the CMB. Highly energetic photon
gases
should exist today in the vicinity of pulsars and magnetars. For
pulsars, the low-frequency dipole radiation is not able to leave the
system directly, as it is first reflected by the surrounding plasma. The
momentum of the low-frequency photons in the vicinity of pulsars are
not necessarily thermally distributed, but still we should be able to
think of the pulsar environment as a highly energetic photon gas. For
magnetars (i.e., pulsars with surface magnetic field strength $\sim
10^{10}\,\mathrm{T}$ \cite{magnetar}), 
the high fields create a surface tension leading
to star quakes, and subsequent generation of low-frequency
photons. This environment could be in an electromagnetically
turbulent state \cite{Kondratyev}. Thus electromagnetic pulses of a
comparatively moderate intensity propagating in such environments may
collapse, due to the high energy densities of the photon gases. Rough
estimates give energy densities of such photon gases that can be
$20-22$ orders of magnitude larger than that of todays CMB.

While the conditions for collapse may be fulfilled only during
extraordinary circumstances, there is still a possibility that
energetic 
pulses, although not intense enough for collapse, can leave a certain 
inprint in the CMB. Since we are mainly interested in effects that
persist 
after the EM-pulse has passed a given area, we need to abandon the
approximation of a broad or needle shaped pulse and instead
investigate solutions to (\ref{Coomoving-acoustic}). 
If the radiation gas energy density is well below that for
collapse, 
the backreaction on the EM-pulse is small, and we do not have to solve
the system (\ref{eq:nlse}) and (\ref{Coomoving-acoustic}) 
self-consistently, but can take the EM-pulse as given. 
Furthermore, investigating the generation of acoustic waves during
times short compared to the diffraction time, we can consider the
pulse to be static in the comoving frame. To simplify the analysis by
making the $r$ and $\xi$-dependence separable, we assume that the
source term obeys  
$ \nabla _{\bot }^{2}\left| E_{p}\right|^{2} + r_0^{-2}\left|
 E_p\right|^2 = 0$,
where $r_0$ is the characteristic radius of localisation. 
For a pulse with no angular dependence, this means that
we can   
write $\left| E_{p}\right| ^{2}=J_{0}(r/r_{0})S(\xi)$, where
$J_{0}$ 
is the $0^{\mathrm{th}}$ order Bessel function, and we leave $S(\xi )$
unspecified. 
Letting the generated acoustic waves have the same radial dependence
as the 
pulse, i.e. $\delta \mathscr{E}(r ,\xi ) =
J_{0}(r/r_{0})\mathscr{E}_{\xi}(\xi)$ we then obtain from 
(\ref{Coomoving-acoustic}) 
\begin{equation}
  \frac{d^2\mathscr{E}_{\xi}}{d\xi^2} +
  \frac{1}{2r_0^2}\mathscr{E}_{\xi} =
  \lambda\varepsilon_0\mathscr{E}_0\left[ r_0^{-2}S -
  2\frac{d^2S}{d\xi^2} \right] .
\label{Acoustic2}
\end{equation}
With the boundary condition of no acoustic waves before the pulse has  
passed, the solution of (\ref{Acoustic2}) is
\begin{equation}
\mathscr{E}_{\xi }(\xi ) = \frac{\sqrt{8}\lambda \varepsilon
_{0}\mathscr{E}_{0}}{r_0} \int_{\infty }^{\xi }S(\xi ^{\prime })\sin
[(\xi -\xi^{\prime })/\sqrt{2}r_0]\,d\xi ^{\prime }. 
\end{equation}
Apparently, after the pulse passage the field can be written as 
$\mathscr{E%
}_{\xi }=\mathscr{E}_{\xi 0}\sin[(\xi/\sqrt{2}r_0) + \delta]$ , where
the 
amplitude $\mathscr{E}_{\xi 0}$ and the phase angle $\delta $ depend
on the 
detailed form of the pulse profile $S(\xi ^{\prime })$ (see
Ref.~\cite{Gorbunov-etal}
for 
analytical expressions). For a pulse length slightly less than $r_0$  
we have $\mathscr{E}_{\xi 0}\sim \lambda \varepsilon _{0}\mathscr{E}%
_{0}S_{\max }\sim \lambda \varepsilon _{0}\mathscr{E}_{0}\left|
E_{p}\right| 
_{\max }^{2}$, where $\left| E_{p}\right| _{\max }^{2}$ is the central
value 
of $\left| E_{p}\right| ^{2}$. 
Currently, measurements of the CMB can detect relative temperature
anisotropies of the order of $10^{-6}$ \cite{wmap}. Thus, we see that
$|E_p| \gtrsim 10^{17}\,\mathrm{V}/\mathrm{m}$ must be fulfilled
for possible detection of the resulting background anisotropy. 
Our estimate for the acoustic wave  
amplitude applies only relatively close after the pulse passage,
before the acoustic wave begins to spread due to diffraction (as
associated with the 
slow $\tau $-dependence omitted in Eq.~(\ref{Coomoving-acoustic})). It
is clear from the above  
that intense pulses leave an inprint in the CMB after the 
pulse has passed a given area, and the results shows that anisotropies
in 
the high-frequency electromagnetic spectrum partly transfer to the
lowfrequency background.  However, our calculations apparently do not
include all effects that will influence the earth-based measurements. 
Thus much work is still needed
in order to determine whether the effects induced in the CMB by
astrophysical sources can be seen in the detailed experiments that
are currently made \cite{wmap}.


\begin{thebibliography}{99}
\bibitem{magnetar} C.\ Kouveliotou \textit{et al.}, Nature
  \textbf{393} 235 (1998). Note that the surface field
    $10^{10}\,\mathrm{T}$ is close to the critical value
    $E_{\text{crit}}/c$, but, because the field strength decreases
    rapidly from the surface and out, the second of the inequalities
    (\ref{eq:constraint}) is also satisfied.
\end{thebibliography}
\end{document}